\documentclass[journal]{IEEEtran}

\usepackage{cite}

\ifCLASSINFOpdf
   \usepackage[pdftex]{graphicx}
   \graphicspath{{../pdf/}{../jpeg/}}
   \DeclareGraphicsExtensions{.pdf,.jpeg,.png}
\else
   \usepackage[dvips]{graphicx}
   \graphicspath{{../eps/}}
   \DeclareGraphicsExtensions{.eps}
\fi

\usepackage{subfig}

\begin{document}
\title{Channel Static Antennas for Mobile Devices}

\author{Gerald Artner,~\IEEEmembership{Member,~IEEE}% <-this % stops a space
\thanks{Gerald Artner was with the Institute of Telecommunication, Technische Universit\"at Wien, Gu\ss hausstra\ss e 25, 1040 Vienna, Austria, e-mail: gerald.artner@nt.tuwien.ac.at, website: geraldartner.com.}% <-this % stops a space
%\thanks{Manuscript received March 8, 2018; revised August 26, 2015.}
}

\markboth{May~2019}%
{Gerald Artner: Channel Static Antennas for Mobile Devices}

\maketitle

\begin{abstract}
Channel static antennas are considered for mobile devices.
The antenna keeps the wireless communication channel static by performing a counter-movement that is opposed to movements of the device that might be caused by a user.
The feasibility of the concept is demonstrated for linear movement in an office environment. 
Channel measurements are performed with quarter wavelength monopole antennas in the $2.4$\,GHz ISM frequency band.
A channel model for the wireless communication channel of mobile devices with channel static antennas is proposed based on these measurement results.
\end{abstract}


%\IEEEpeerreviewmaketitle

\section{Introduction}

It has recently been proposed, that devices can keep their wireless communication channels static while moving or being moved by performing a counter-movement of the antenna \cite{ArtnerCSA,ArtnerCSAExperiment}.
However, small devices for mobile usage are limited in their size and therefore limited in the distance over which they can keep their channels static with a counter-movement.
Among these devices are prominent use cases such as smart phones, tablets, laptops, pagers, IoT devices, virtual reality headsets etc.

\emph{Contribution} --- A first investigation of channel static antennas for mobile devices is conducted experimentally.
It is considered that antennas on such devices are able to keep wireless communication channels static by performing counter-movements within the limitations of the device's size.
Measurements were performed in a laboratory/office environment with quarter-wavelength monopole antennas at $2.45$\,GHz.
The antennas were measured alone for this first proof of concept -- channel static antennas were not prototyped as part of a mobile device.
Channel models are proposed for mobile devices with channel static antennas. 

\section{Theoretical Considerations}
\label{sec_theory}

\begin{figure}[!t]
\centering
\subfloat[]{\includegraphics[width=0.15\textwidth]{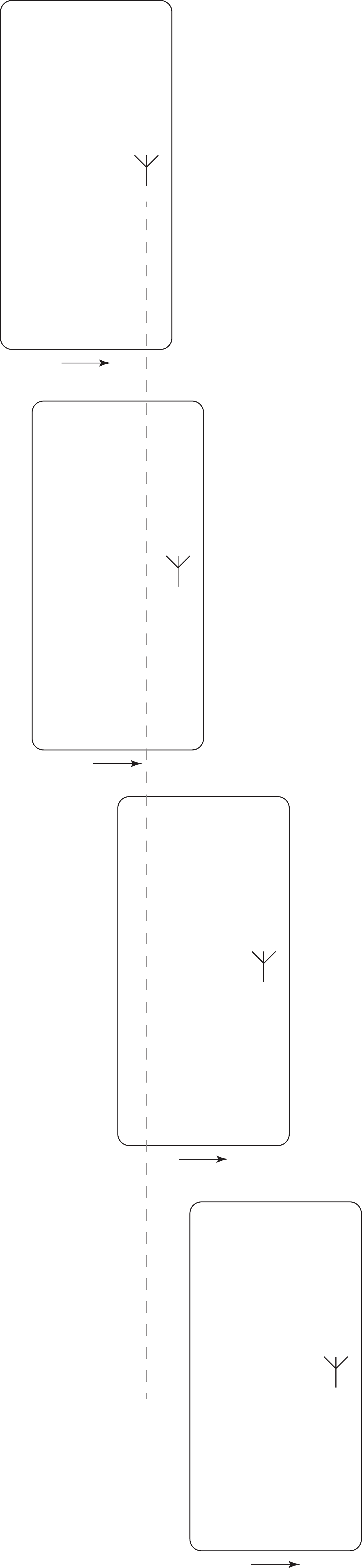}
\label{fig_phone_regular}}
~
\subfloat[]{\includegraphics[width=0.15\textwidth]{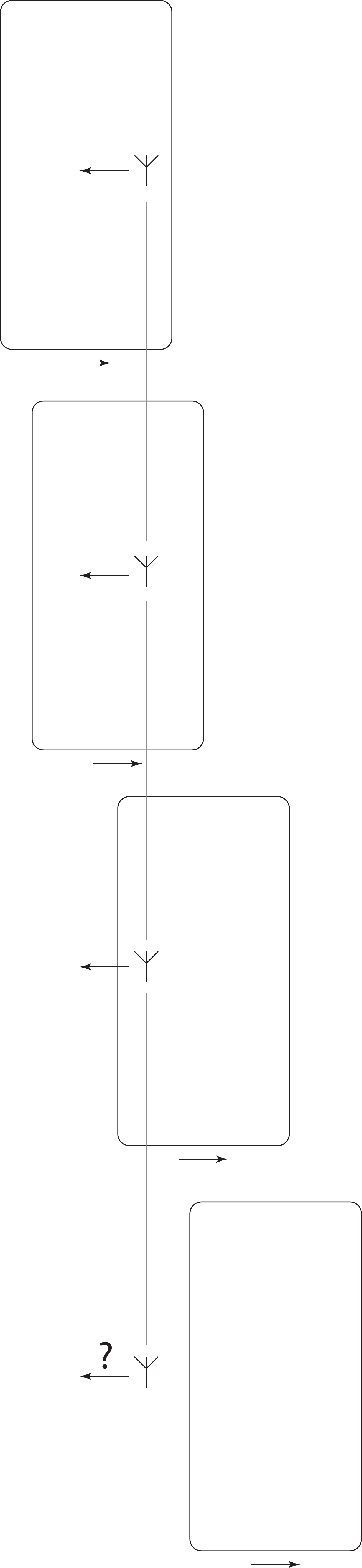}
\label{fig_phone_CSA}}
~
\subfloat[]{\includegraphics[width=0.15\textwidth]{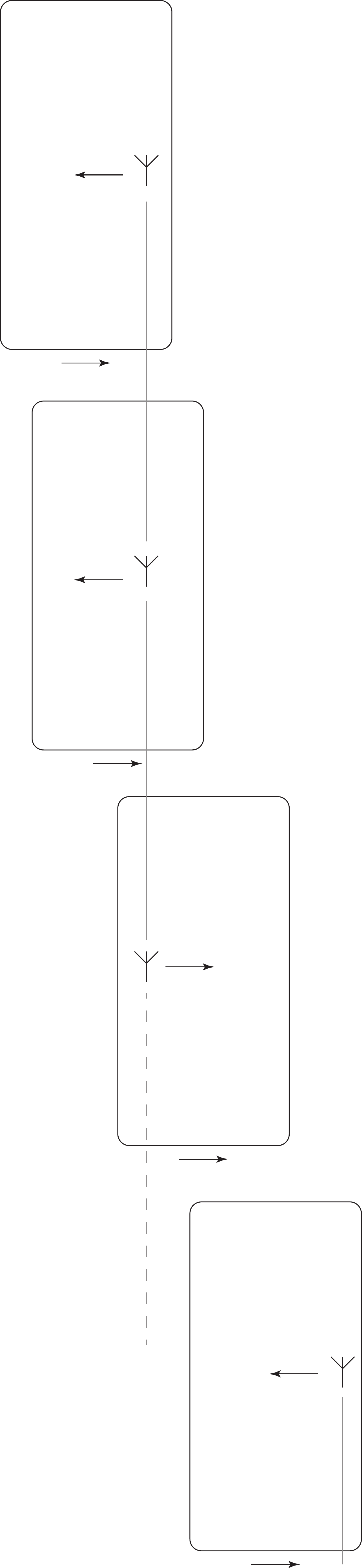}
\label{fig_phone_CSAH}}
\caption{a) A state of the art mobile device. The device moves to the right. The antenna is fixed to the device and moves with it. The communication channel will change because the antenna moved away from its original position
b) Channel static antenna \cite{ArtnerCSAExperiment}: The device moves to the right. The antenna performs a counter-movement to stay in its original position. This keeps the wireless channel static, because the antenna has not moved from the viewpoint of an outside observer. However, the antenna can not move beyond the size limitations of the device (bottom).
c) A channel static antenna for mobile devices is investigated in this work. The antenna again performs a counter-movement. It obeys the size limitations of its device. When it reaches the end, it moves to a different position on the device. At the new position a new channel is formed, which the antenna keeps again static and so on. The channel static antenna for the mobile device therefore keeps the channel \emph{piecewise} static.}
\label{fig_sketch_phone}
\end{figure}

Movement of devices generally changes the wireless communication channels that these devices experience, if the antennas are fixed to the device (which is the state of the art today).
In \cite{ArtnerCSA,ArtnerCSAExperiment} it is considered that the antenna can keep the channel static by performing a movement on the device which is in opposite direction of the device's movement (counter-movement).
From the viewpoint of an outside observer, the counter-movement keeps the antenna in its original position.
This in turn keeps the channel static towards outside observers such as other antennas at arbitrary locations.

The counter-movement of the antenna is of course limited by the size of its device.
To avoid falling off, the antenna backs off and chooses a new position on the device.
There, the antenna again performs counter-movements to stay in that position to keep the channel static until it reaches a device edge and so on.
Therefore, the channel static antenna for mobile devices keeps the channel \emph{piecewise} static for long device movements.
The principle is sketched in Fig.~\ref{fig_sketch_phone}.

In practice it can be expected that a counter-movement of the antenna on a small device will influence the radiation characteristics of the antenna, e.g. the antenna position has a significant influence on cars \cite{Jesch1985,Yang2018}.
Such influences are not considered within the scope of this work.
It will be the pleasure of the device's designer to keep the antenna characteristics within given bounds during the counter-movement.

\section{Experimental Proof of Concept}
\label{sec_experiment}

\begin{figure}[!t]
\centering
\includegraphics[width=0.49\textwidth]{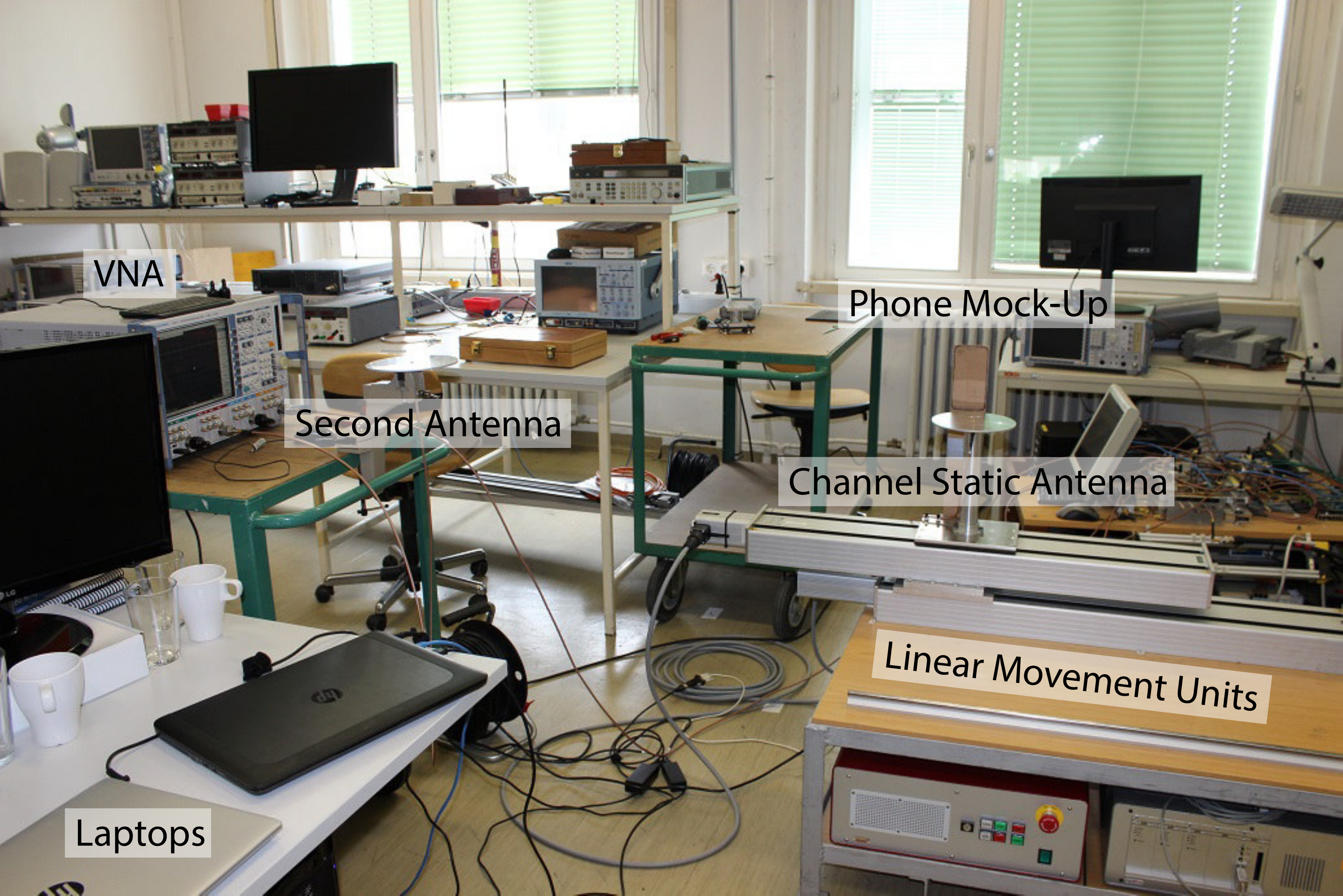}
\caption{The measurement setup and the office environment.}
\label{fig_office}
\end{figure}

The concept of channel static antennas is experimentally demonstrated for handheld devices in an office and laboratory environment.
Without loss of generality, the device trajectory is confined to linear movement over a finite distance.
Two quarter-wavelength monopole antennas for the $2.4$\,GHz ISM frequency band are placed on circular aluminum ground planes.
They are connected to a vector network analyzer (VNA) to measure the channel between them in absolute value and phase.
The antennas are placed a few meters apart ($120$\,cm at the intial position) in an office room with laboratory equipment.
From an electromagnetic viewpoint, this is a complex environment with different geometries and materials.
It is also a typical environment for WLAN operation in the $2.4$\,GHz band.
One antenna is fixed in its position.
The other antenna is mounted on two stacked linear movement units.
The bottom unit moves the top unit.
The top unit performs the counter-movement of the antenna.
The linear movement units and the VNA are connected to a laptop, which controls the experiment.
The VNA measurements take some time.
During this time the antenna remains still, otherwise the movement would influence the results.
As a consequence, no Doppler shift is present in the channel.
Fig.~\ref{fig_office} shows the office environment and the experimental setup.

\begin{figure}[!t]
\centering
\includegraphics[width=0.49\textwidth]{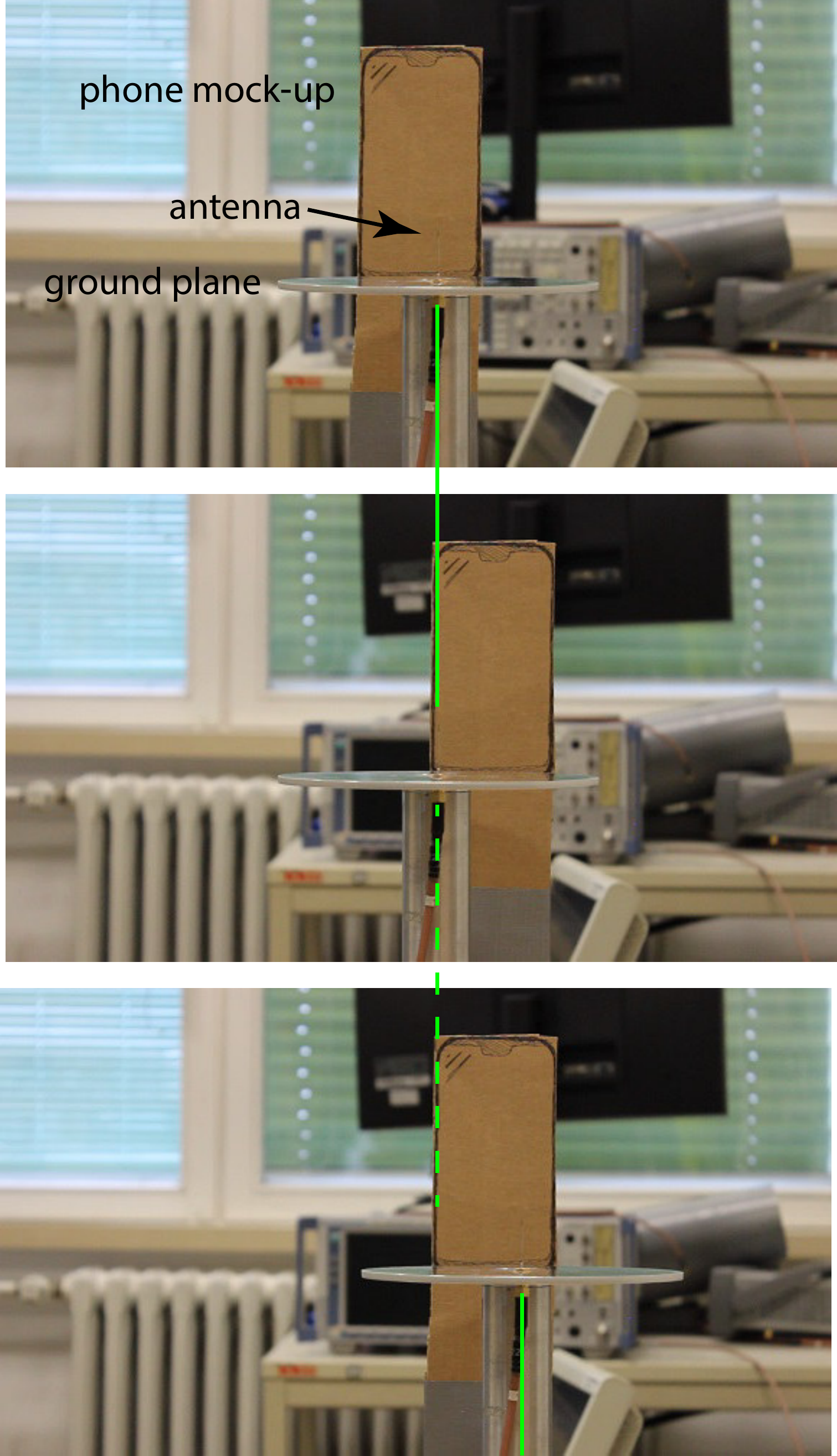}
\caption{The antenna keeps the channel static within the size limits of a mobile device. Top to middle: The device moves to the right. The antenna will perform counter-movements to stay in its position relative to outside observers. Middle to bottom: The antenna can not move beyond the device. It moves to a new position at which it keeps the channel static.}
\label{fig_phone}
\end{figure}

The distance of the counter-movement is limited to $0.5$\,$\lambda$ (about $6$\,cm).
This is a feasible distance considering the width of todays smart phones \cite{SmartPhoneSizes}.
A thin cardboard smart phone mockup is added behind the antenna as a visual reference.
Its influence is considered to be negligible.
The device mock-up is moved over a distance $6$\,$\lambda$ that far exceeds its width.
Photographs that show illustrate the working principle are shown in Fig.~\ref{fig_phone}.

Three measurements are performed.
First, the antenna moves away from its initial position with a linear motion.
This is the case with state of the art antennas that are fixed to mobile devices.
Second, the antenna performs counter-movements to keep the channel static under this linear motion.
Third, the antenna remains still in its initial position.
This serves as a reference to see how static the channel remains without movement, i.e. to quantify channel changes that are caused by the environment.

\begin{figure}[!t]
\centering
\subfloat[]{\includegraphics[width=0.49\textwidth]{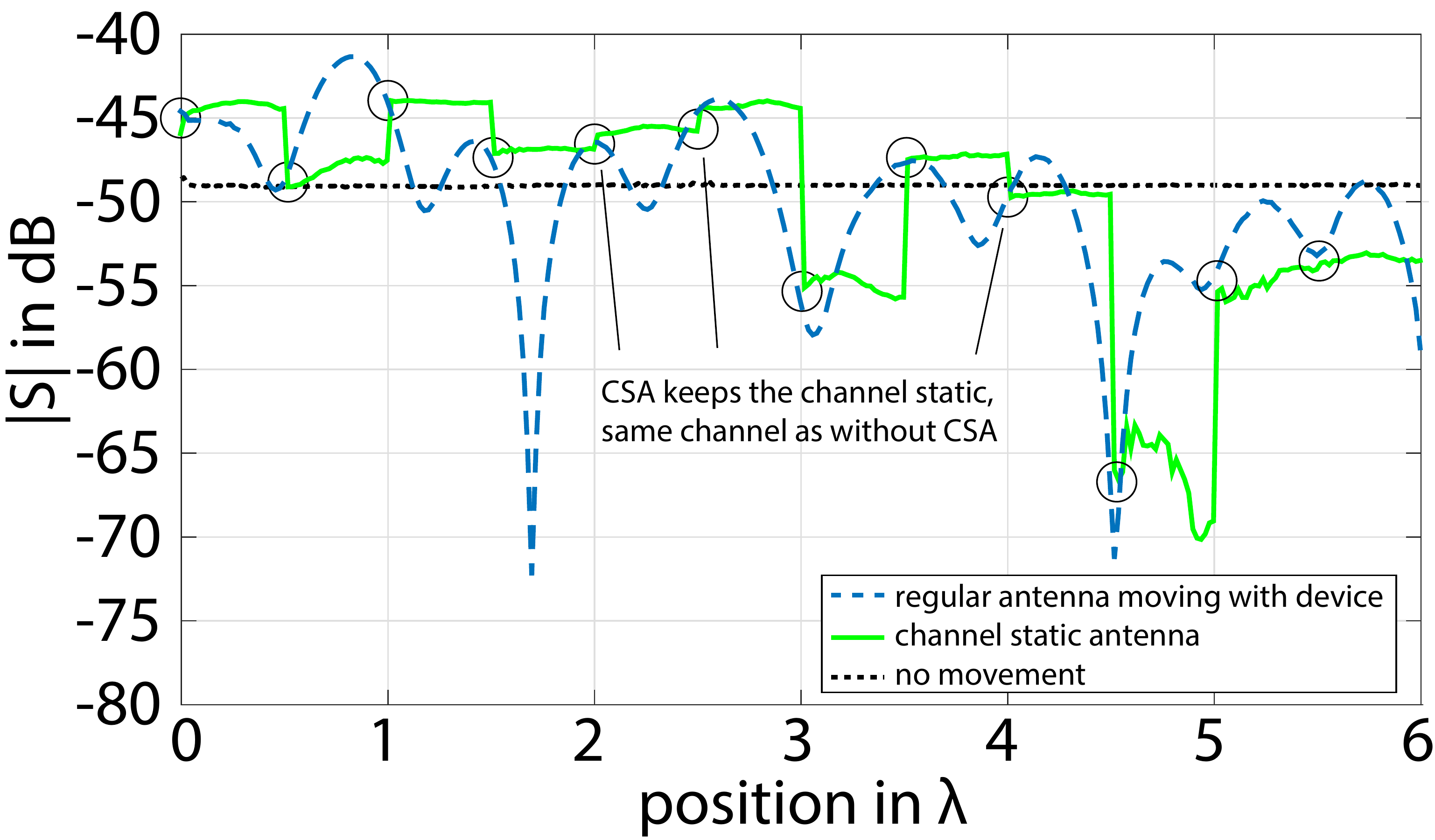}
\label{fig_results_amp}}
\hfil
\subfloat[]{\includegraphics[width=0.49\textwidth]{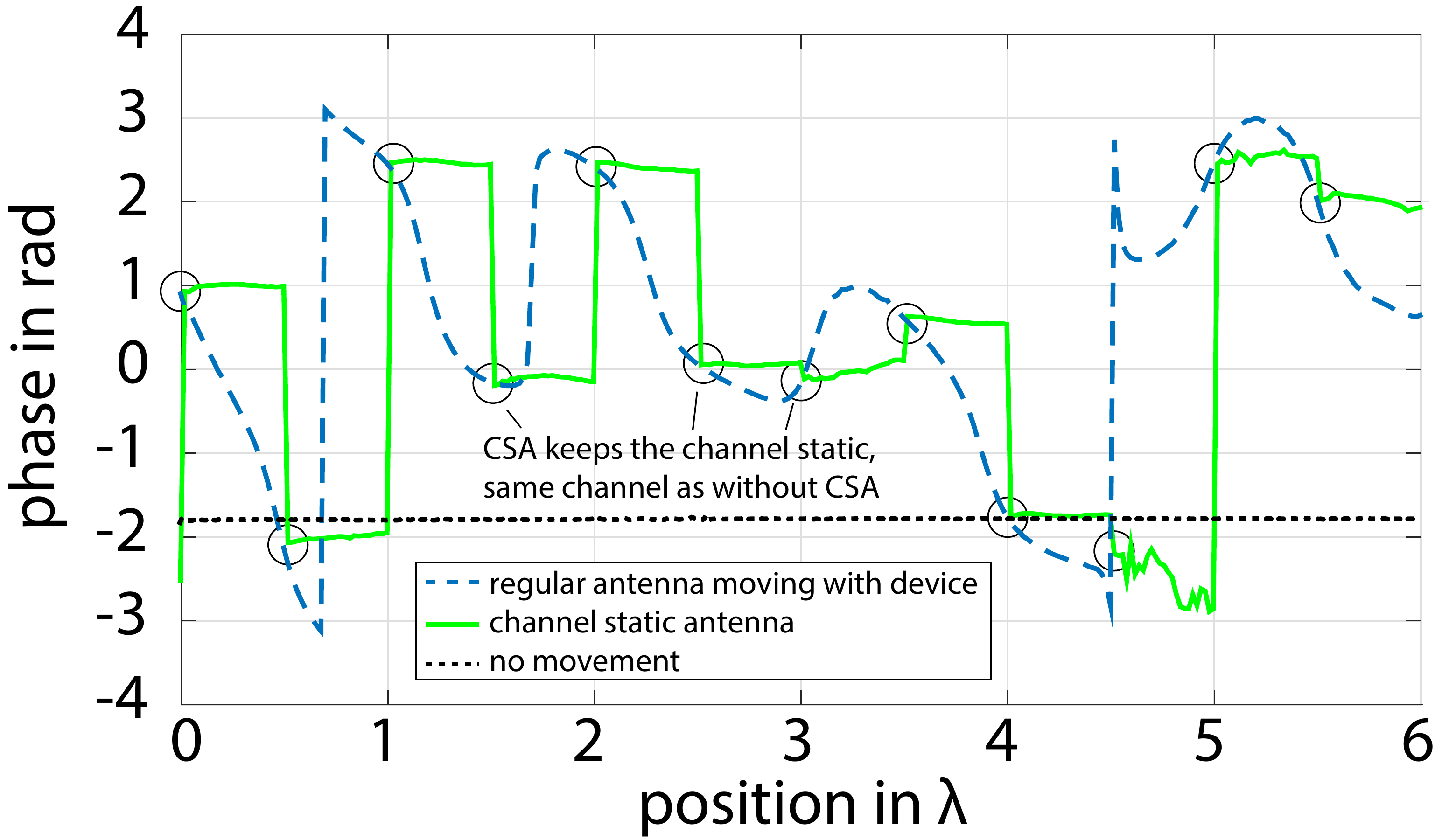}
\label{fig_results_phase}}
\caption{Measurement results of the CSA under size limitations a) amplitude and b) phase. The phase is wrapped at $2\pi$ for convenient viewing.}
\label{fig_results}
\end{figure}

The measured amplitude and phase of the channel are shown in Fig.~\ref{fig_results}.
The results without compensation and with the channel static antenna are plotted as a function of the distance that the mobile device has moved from the initial position in wavelengths.
The results without antenna movement are plotted for a similar time period as the other measurements took to provide a comparison.

The results show that wireless communication channels can be kept \emph{piecewise} static under device movement by performing a counter-movement of the antenna within the size limitations of the device.
The channel is kept static in both amplitude and phase.
A regular moving antenna moves through the small scale fading environment that is typical of indoor office environments.
The antenna experiences large variations in the channel and deep fading notches with $20$\,dB reduced receive power.
With the channel static antenna the channel stays practically static until the antenna reaches the end of the device and has to move to a new channel.
The initial static channel coincides with the channel that a regular antenna would experience at this position.
The channel static antenna keeps the wireless channel static, but does not ensure, that it is a good channel, e.g. the channel at distance $4.5$\,$\lambda$ coincides with a fading notch and this position is kept although the channel would get better.

There are small residual changes during a static interval.
This variation is a lot smaller in the measurement without movement, which hints that the residual changes stem from the process that keeps the channel static and not from environmental changes.
The influences might come from the cardboard mock-up, but they are more likely caused by the large metallic movement unit.
Overall, the channel static antenna does an excellent job in keeping the channel static when compared to regular antennas.
A device equipped with channel static antennas would no longer experience small scale fading or fast fading when moving through the environment.

\section{A Channel Model for Channel Static Antennas}
\label{sec_model}

The measurements in Sec.~\ref{sec_experiment} suggest that the channel is indeed kept static when the device moves.
The channel at subsequent positions of the device continues to be channel from the initial position where the CSA started to keep the channel static.
The channel remains until the antenna can no longer keep it static due to technological constraints.
In the investigated scheme the antenna is constrained in its counter-movement by the size of the device.
Other schemes that aim to keep channels static might have different limitations \cite{ArtnerPCSA,ArtnerPCSAExperiment}.
Based on the measurements in Sec.~\ref{sec_experiment} the following channel model is proposed.

\begin{equation}
H(n) = H(n_0)
\label{eq_CSA}
\end{equation}

where $H(n)$ is the channel at device distance $n$ from an initial position and $H(n_0)$ is the channel at the intial position $n_0$ where the channel is kept static.

Note that the model does not dispose of the initial channel $H(n_0)$.
$H(n_0)$ can be the real channel that forms in the environment, it can be an estimate of the real channel \cite{Simko2013}, it can be obtained from a channel model \cite{WinnerII,COST2100} or drawn from a distribution \cite{Zochmann2019}.

Instead of a spatial formulation, the model can be formulated in the temporal domain as

\begin{equation}
H(t) = H(t_0)
\label{eq_CSA_t}
\end{equation}
where $H(t)$ is the channel at time $t$ and $H(t_0)$ is the channel at time $t_0$ when the antenna started to keep the channel static.

In many applications it will be suitable to modify the initial channel $H(n_0)$ with some function $F$:
\begin{equation}
H(n) = F\left(H(n_0)\right)
\label{eq_CSA_F}
\end{equation}
The function $F$ might add noise, consider estimation uncertainty or model a technological process that is used to keep the channel static, e.g. the influence of moving a device in the near-field of an antenna.
Exemplary, for an office environment (as in Sec.~\ref{sec_experiment}) the channel might be modeled as
\begin{equation}
H(n) = H(n_0) + N,
\label{eq_CSA_office}
\end{equation}
where $H(n_0)$ might be a Rice-distributed random variable that is drawn at new initial positions $n_0$ and stays constant during a static interval.
It models the small scale fading environment of the office.
$N$ might be a Gau\ss -distributed noise term that is drawn at each position $n$ and models the residual channel changes during a static interval.
%The specific measurement in Fig.~\ref{fig_results} might then be modeled as:
%\begin{equation}
%H(n) = H(n_0) + N,
%\label{eq_CSA_office_fit}
%\end{equation}
%with $H(n_0) \sim \textrm{Rice}\left(3.64 \cdot 10^{-3} , 2.96 \cdot 10^{-6}\right)$ and $N \sim \textrm{Gau\ss}\left( 0, 2 \cdot 10^{-9} \right)$

Antenna movement over large distances can cause changes in the channel statistics, e.g. wide-sense stationary assumptions no longer hold for vehicular channels.
For characterization, such channels can be divided into smaller chunks during which the channel statics do not change significantly \cite{Matz2005}.
Such procedures are widely used, but the division into channel pieces is somewhat arbitrary.
When channels are kept static with the proposed procedure, then it can be expected that the variations within a static interval are quite a lot smaller than when they are allowed to change freely (e.g. with Rayleigh or Ricean fading) -- as is demonstrated by the measurements in Sec.~\ref{sec_experiment}.
The author expects that such models might be well suited for channel static antennas, as the statistics of the initial channels $H(n)$ might change, but the statistics of the residual fluctuations remain the same within a static interval.
The division of the whole channel into chunks is then no longer arbitrary, but linked to the intervals where the channel is kept static with the channel static antenna.

\section{Conclusion}

The proposed technique keeps the wireless communication channel static under device movement by performing a counter-movement of the antenna that considers the limited size of mobile devices.
Feasibility of the concept is shown with an experiment in an office environment.
The performance of channel static antennas for mobile devices is measured and assessed.
In such applications, channel static antennas are able to keep the channel \emph{piecewise} static.
The technique is purely based on the device's movement and size, the antenna does not require channel knowledge to perform the counter-movement. 
Modern smart phones are already equipped with sensors that measure the position, acceleration and tilting of devices.

First mathematical models are presented to describe wireless communication channels that are kept static by channel static antennas.

\section*{Acknowledgment}

The author thanks M.~Lerch of Technische Universit\"at Wien, Vienna, Austria, for his help with the experimental work.

\ifCLASSOPTIONcaptionsoff
  \newpage
\fi



\begin{thebibliography}{1}

%\bibitem{Artner2014SAM}
%G.~Artner, M.~Mayer, M.~Guillaud and M.~Rupp, ``Measuring the Impact of Outdated Channel State Information in Interference Alignment Techniques,'' \emph{Proc. IEEE 8th Sensor Array and Multichannel Signal Processing Workshop (SAM)}, pp.353-356, 2014.
%
%\bibitem{Saligheh2008}
%H.~Saligheh~Rad and S.~Gazor, ``Effects of mobile rotational movements in wireless propagation channels,'' \emph{IET Communications}, vol. 2, no. 9, 2008, pp.1109-1117.
%
%\bibitem{Mecklenbrauker2011}
%C.F.~Mecklenbr\"auker, A.F.~Molisch, J.~Karedal, F.~Tufvesson, A.~Paier, L.~Bernad\'o, T.~Zemen, O.~Klemp and N.~Czink, ``Vehicular Channel Characterization and Its Implications for Wireless System  Design and Performance,'' \emph{Proceedings of the IEEE}, vol. 99, no. 7, 2011, pp.1189-1212.
%
%\bibitem{Yang2018}
%J.~Yang, B.~Ai, K.~Guan, D.~He, X.~Lin, B.~Hui, J.~Kim, A.~Hrovat , ``A Geometry-Based Stochastic Channel Model for the Millimeter-Wave Band in a 3GPP High-Speed Train Scenario,'' \emph{IEEE Transactions on Vehicular Technology}, vol. 67, no. 5, 2018, pp.3853-3865.

%\bibitem{Artner2013Dipl}
%G.~Artner, ``Receiver Location Sensitivity of Interference Alignment,'' Diploma Thesis, Technischen Universit\"at Wien, Institute of Telecommunications, Vienna, September 2013.
%
%\bibitem{Zochmann2019}
%E.~Z\"ochmann, S.~Caban, C.F.~Mecklenbr\"auker, S.~Pratschner, M.~Lerch, S.~Schwarz and M.~Rupp:, ``Better than Rician: modelling millimetre wave channels as two-wave with diffuse power,'' \emph{EURASIP Journal on Wireless Communications and Networking}, vol. 2019, no. 1, 2019, pp.1-17.
%
%\bibitem{Simko2013}
%M.~Simko, P.S.R.~Diniz, Q.~Wang and M.~Rupp, ``Adaptive Pilot-Symbol Patterns for MIMO OFDM Systems,'' \emph{IEEE Trans. Wirel. Comm.}, vol. 12, no. 9, 2013, pp.4705-4715.
%
%\bibitem{Engiz2015}
%B.K.~Engiz, C.~Kurnaz and H.~Sezgin, ``Approach for determining the optimum pilot placement in orthogonal frequency division multiplexing systems,'' \emph{IET Communications}, vol. 9, no. 15, 2015, pp.1915-1923.
%
%\bibitem{Sternad2012}
%M.~Sternad, M.~Grieger, R.~Apelfr\"ojd, T.~Svensson, D.~Aronsson and A.B.~Martinez, ``Using ``predictor antennas'' for long-range prediction of fast fading for moving relays,'' in Proc. \emph{IEEE  Wireless Communications and Networking Conference Workshops (WCNCW)}, 2012, pp.253-257.
%
%\bibitem{Bjorsell2017} J. Bj\"orsell, M.~Sternad and M.~Grieger, ``Using predictor antennas for the prediction of small-scale fading provides an order-of-magnitude improvement of prediction horizons,''  in Proc. \emph{International Conference on Communications (ICC)}, 2017.
%
%\bibitem{Jaeck2017}
%V.~Jaeck, L.~Bernard, K.~Mahdjoubi, R.~Sauleau, S.~Collardey, P.~Pouliguen and P.~Potier, ``A Switched-Beam Conformal Array With a 3-D Beam Forming Capability in C-Band,'' \emph{IEEE Transactions on Antennas and Propagation}, vol. 65, no. 6, 2017, pp.2950-2057.
%
%\bibitem{Hanna1947}
%C.R.~Hanna and L.B.~Lynn, ``Gyroscope Controlled Antenna Stabilizer,'' US Patent 2,425,737, 1947.
%
%\bibitem{Goss1955}
%F.A.~Goss~Jr., ``Mechanical Stabilizer for Supporting Radar Antenna,'' US Patent 2,706,781, 1955.
%
%\bibitem{Besso2007}
%P.~Besso, M.~Bozzi, M.~Formaggi and L.~Perregrini, ``A Novel Technique for High-Performance Correction of Beam Aberration in Deep Space Antennas,'' \emph{IEEE Ant. Wirel. Propag. Lett.} vol. 6, 2007, pp.376-378.

\bibitem{ArtnerCSA}
G.~Artner, ``Antennenanordnung f\"ur statische Funkkan\"ale,'' ̈\"Osterreichische Patentanmeldung A51035/2018, 2018.

\bibitem{ArtnerCSAExperiment}
G.~Artner, ``Channel Static Antennas,'' arXiv:1904.01627v1, 29 Mar 2019.

\bibitem{Jesch1985}
R.L.~Jesch, ``Measured Vehicular Antenna Performance,'' \emph{IEEE Trans. Veh. Techn.}, vol. VT-34, No. 2, 1985, pp. 97-107.

\bibitem{Yang2018}
J.~Yang, J.~Li and S.~Zhou, ``Study of Antenna Position on Vehicle by Using a Characteristic Modes Theory,'' \emph{IEEE Antennas and Wireless Propagation Letters}, vol. 17, no. 7, 2018, pp.1132-1135.

\bibitem{SmartPhoneSizes}
https://mobiledevicesize.com

\bibitem{Simko2013}
M. \v{S}imko, P.S.R. Diniz, Q. Wang, M. Rupp, ``Adaptive pilot-symbol patterns for MIMO OFDM systems,'' \emph{IEEE Transactions on Wireless Communications}, vol. 12, no. 9, 2013, pp.4705-4715.

\bibitem{WinnerII}
P.~Ky\"osti, J.~Meinil\"a, L.~Hentil\"a, X.~Zhao, T.~J\"ams\"a, C.~Schneider, M.~Narandzi\'c, M.~Milojevi\'c, A.~Hong, J.~Ylitalo, V.-M.~Holappa, M.~Alatossava, R.~Bultitude, Y.~de~Jong and T.~Rautiainen, ``WINNER II Channel Models,'' 2007.

\bibitem{COST2100}
L.~Liu, C.~Oestges, J.~Poutanen, K.~Haneda, P.~Vainikainen, F.~Quitin, F.~Tufvesson and P.~de~Doncker, ``The COST 2100 MIMO channel model,'' \emph{IEEE Wireless Communications}, vol. 19, no. 6, 2012, pp.92-99.

\bibitem{Zochmann2019}
E.~Z\"ochmann, S.~Caban, C.F.~Mecklenbr\"auker, S.~Pratschner, M.~Lerch, S.~Schwarz and M.~Rupp, ``Better than Rician: modelling millimetre wave channels as two-wave with diffuse power,'' \emph{EURASIP Journal on Wireless Communications and Networking}, vol. 1, 2019.

\bibitem{Matz2005}
G.~Matz, ``On non-WSSUS wireless fading channels,'' \emph{IEEE Trans. Wireless Commun.}, vol. 4, no. 5, 2005, pp. 2465-2478.


%\bibitem{Doppler1842}
%C.~Doppler, ``Ueber das farbige Licht der Doppelsterne,'' Borrosch \& Andr\'e, Prag, 1842.
%
\bibitem{ArtnerPCSA}
G.~Artner, ``Kanalstatische Antenne zur Kompensation von Bewegungen einer Partnerantenne,'' \"Osterreichische Provisorische Patentanmeldung A60105/2019, 2019.

\bibitem{ArtnerPCSAExperiment}
G.~Artner, `´Channel Static Antennas for Compensating the Movements of a Partner Antenna,'' unpublished manuscript.

\end{thebibliography}
\end{document}